\documentstyle[]{l-school}
\input psfig
\newcommand{\beq}{\begin{equation}}
\newcommand{\eeq}{\end{equation}}
\newcommand{\bea}{\begin{eqnarray}}
\newcommand{\eea}{\end{eqnarray}}
\newcommand{\un}[1]{{\bf #1}}
\newcommand{\half}{{\scriptstyle{{1\over 2}}}}
\newcommand{\real}{\relax{\rm I\kern-.18em R}}
\newcommand{\ad}{{\rm ad}}
\newcommand{\id}{\mbox{$id$}}
\newcommand{\norm}[1]{\left\| #1 \right\|}
\newcommand{\sgbar}{\sg^\dagger}
\newcommand{\tr}{\mbox{\,tr\,}}
\newcommand{\al}{\alpha}
\newcommand{\Gm}{\Gamma}
\newcommand{\Lm}{\Lambda}
\newcommand{\sg}{\sigma}
\newcommand{\Om}{\Omega}
\newcommand{\Ss}[1]{\mbox{$\cal #1$}}
\newcommand{\pr}{\partial}
\newcommand{\Order}[1]{\Ss{O}\left(#1\right)}
\begin{document}
% Preprint numbers
\hfill NI97025NQF

\hfill INLO-PUB-4/97
\vskip-1.2cm
\markboth{Pierre van Baal}{Intermediate Volumes}
\setcounter{part}{1}
\title{Intermediate Volumes and the Role of Instantons}
\author{Pierre van Baal}
\institute{Isaac Newton Institute for Mathematical Sciences,\\ 
20 Clarkson Road, Cambridge CB3 0EH, UK\\and\\
{}\footnote{Permanent address.
Talk given at the workshop "New non-perterturbative methods and quantization 
on the light cone", Les Houches, 24 Feb-7 March, 1997.
}~Instituut-Lorentz for Theoretical Physics,\\
University of Leiden, P.O.Box 9506,\\ NL-2300 RA Leiden, The Netherlands}
\maketitle
\section{Introduction}
An outstanding problem is to understand the formation of a mass gap and 
the spectrum of excitations in a non-Abelian gauge theory. Non-perturbative
aspects are believed to play a crucial role, but despite much progress a 
simple explanation is still lacking. Over the years we have been interested
in addressing this problem in a finite volume, where its size can be used
as a control parameter, which is conspicuously absent in infinite volumes, 
in particular for formulating the binding of gluons in glueballs. Much 
progress was made in intermediate volumes with a torodial geometry, where 
results can be directly compared to lattice Monte Carlo calculations in the 
same physical volume~\cite{vba1}.

The essential features of this analysis are easily explained. At very small
volumes the effective coupling constant is small, due to asymptotic freedom
of non-Abelian gauge theories. In this domain ordinary perturbation theory
can be used. For a torus, due to the presence of zero-momentum modes, for 
which the classical potential is quartic, this results in an expansion in 
powers of $g^{2/3}$ for the spectrum~\cite{lues}. In a spherical geometry, 
where due to curvature of the manifold no zero-modes appear, perturbation 
theory is as usual~\cite{vbda}.

\section{The Role of Instantons}
Irrespective of the geometry of the space on which the gauge theory is
formulated there are low-energy modes in terms of which the wave functional
at larger coupling (i.e. larger volume) will start to spread out over field
space. Not only is the physical Yang-Mills field space a curved 
manifold~\cite{bavi}, but also it has non-trivial topology~\cite{sing}.
In particular the latter is crucial for a better understanding of the 
non-perturbative dynamics. As an example, consider the instantons in the 
Hamiltonian formulation of the theory. They correspond to a path in field 
space associated with minimal action. The stability of the instanton is
guaranteed because the path interpolates between vacua related by a 
topologically non-trivial gauge transformation. This guarantees that the path 
has non-trivial homotopy. Given the non-trivial action, there exists a 
non-zero potential barrier of minimal height which is called a sphaleron and 
exists because the size of the instantons is restricted by the size of the 
volume. It is the energy of this sphaleron that sets the scale beyond which 
the wave functional is no longer exponentially suppressed below the barriers 
separating different vacua. If this is the case, it is no longer possible to 
take instantons into account semiclassically. In essence, instanton solutions 
are used to find the relevant degrees of freedom in the Yang-Mills 
configuration space in whose directions the wave functional will 
first and foremost spread out.

\section{Boundary Conditions in Field Space}
One way of formulating the gauge field configuration space is to use a simple
gauge condition as a parametrisation. Locally it is easily shown that this 
provides a unique description, but since the work of Gribov~\cite{grib} one 
knows that such gauge conditions do not uniquely fix the gauge when moving 
away from the origin in field space. Using a background field gauge fixing, 
one can in principle cover field space by local neighbourhoods, with transition 
functions relating the different neighbourhoods~\cite{nahm}. These transition 
functions are gauge transformations relating gauges of overlapping 
patches~\cite{vba2}. Because field space is infinite dimensional, except for 
low-dimensional models, no satisfactory theory has been developed along these 
lines. Instead we introduce complete gauge fixing using a variational 
formulation of the Coulomb gauge~\cite{sefr}, as in this gauge the Yang-Mills 
Hamiltonian has been studied extensively~\cite{chle}. Minimising the $L^2$ norm
of the vector potential, $A_i(x)=iA^a_i(x)\tau_a/2$, along the gauge orbit
\beq
\norm{^h A}^2~=~ -\int_M d^3x~
\tr \left( \left( h^{-1} A_i h + h^{-1} \pr_i h \right)^2\right),
\label{gAnorm}
\end{equation}
one {\em almost} uniquely fixes the gauge. 
Expanding around the minimum $A$ using $h(x)=\exp(X(x))$, one finds:
\beq
\norm{^h A}^2 = \norm{A}^2+2\int_M \tr(X
\partial_i A_i)+\int_M \tr (X^\dagger FP (A) X)+\Order{X^3},
\label{Xexpansie}
\eeq
where $FP(A)=-\partial_i D_i (A)=-\partial^2_i-\partial_i\ad(A_i)$ is the 
Faddeev-Popov operator $(\ad(A)X~\equiv~[A,X])$. At any local minimum the 
vector potential is therefore transverse, $\partial_i A_i~=~0$, and $FP(A)$ 
is a positive operator. The set of all these vector potentials is by definition
the Gribov region $\Omega$. Using the fact that $FP(A)$ is linear in $A$, 
$\Omega$ is seen to be a convex subspace of the set of transverse connections 
$\Gamma$. Its boundary $\partial \Omega$ is called the Gribov horizon. At the 
Gribov horizon, the {\em lowest} non-trivial eigenvalue of the Faddeev-Popov 
operator vanishes, and points on $\partial\Omega$ are associated with 
coordinate singularities, which can be shown to have a finite distance 
to the origin of field space~\cite{dezw}. 

\begin{figure}{\tt}
\vspace{5.7cm}
\includegraphics{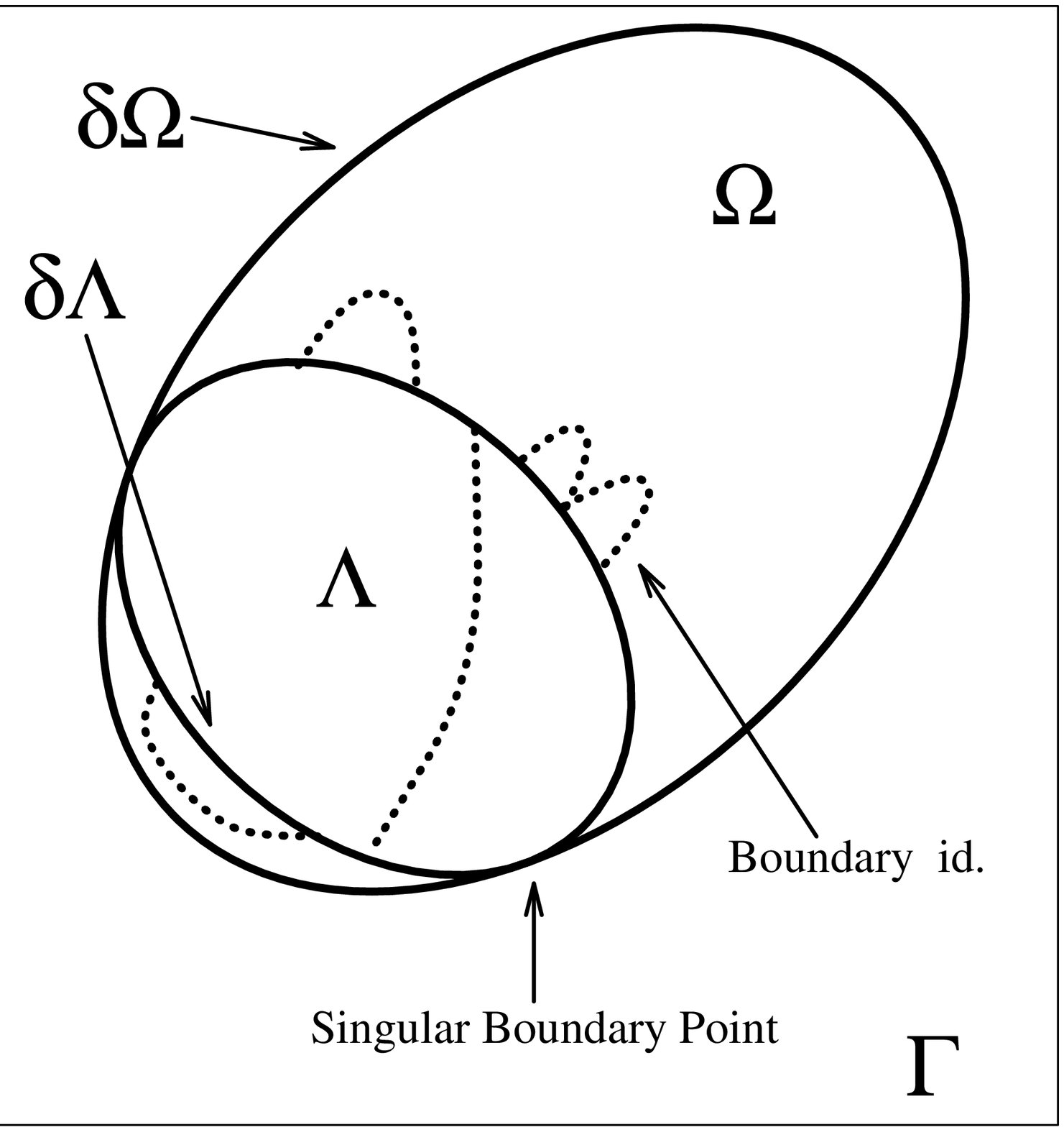}
\includegraphics{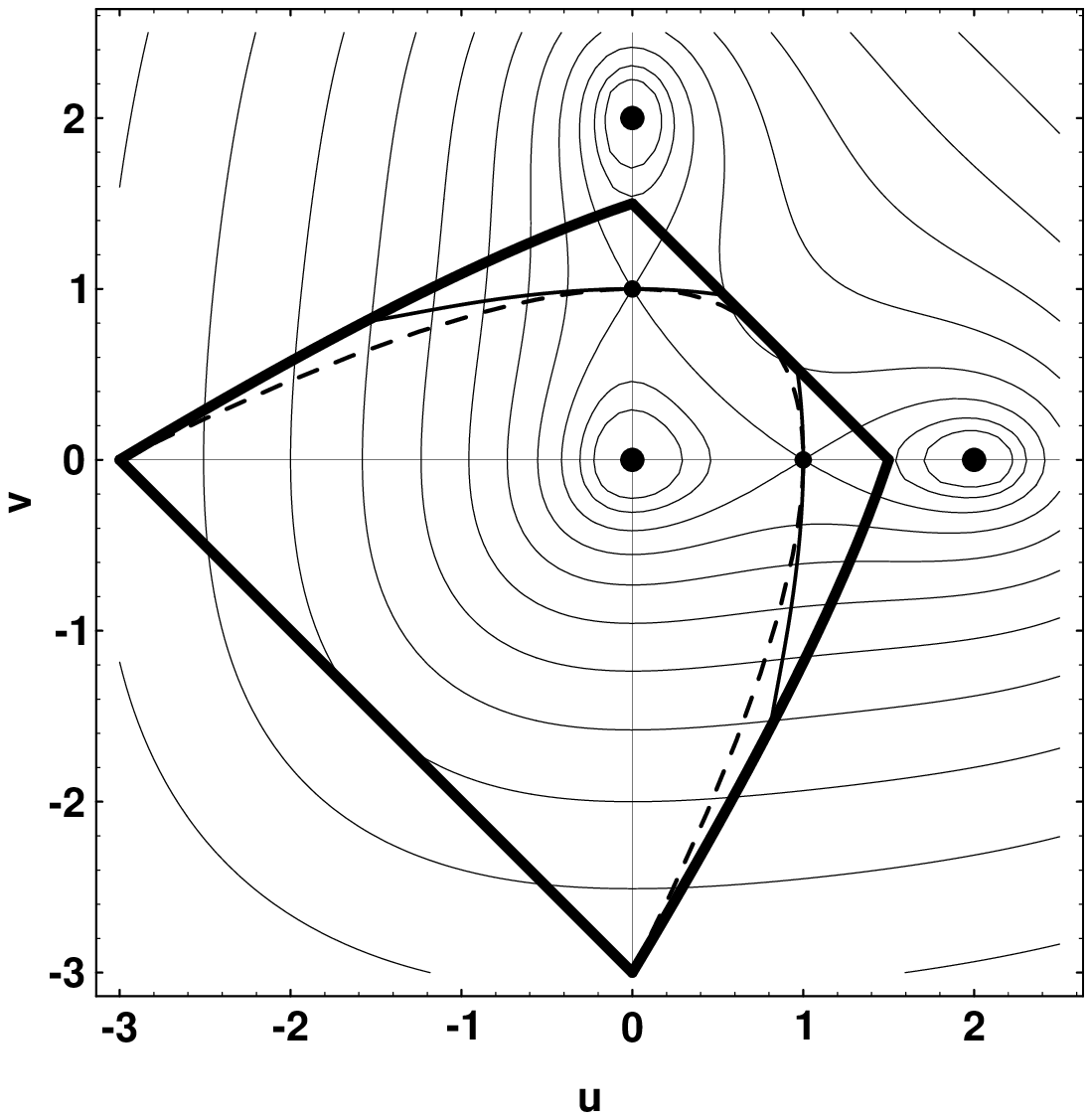}
\caption{On the right a sketch of the fundamental and Gribov regions. The
dotted lines indicate the boundary identifications. On the left a two
dimensional cross section through the configuration space for $S^3$. Location
of the classical vacua (large dots), sphalerons (smaller dots), the Gribov
horizon (fat sections), the horizon for $FP_f$ (dashed curves) which is
contained in the boundary of the fundamental domain (full curves). Also
indicated are the lines of equal potential in units of $2^n$ times the
sphaleron energy.}
\label{fig:fig1}
\end{figure}

The Gribov region, formed by the {\em local} minima, needs to be further 
restricted to the {\em absolute} minima to form a fundamental domain, denoted 
by $\Lambda$. One expects many relative minima, as local gauge functions 
$h(x)\in\Ss{G}$ are like spin variables, noting similarity to the spin glass 
problem. We can write
\beq
  \Lm=\left\{A\in\Gm|\min_{h\in\Ss{G}}\left[\norm{^hA}^2-\norm{A}^2=\int\tr
  \left(h^\dagger FP_f(A)~h\right)\right]=0\right\},\label{FPhalfdef}
\eeq
where $FP_f(A)=-\pr_i^2-i A^a_i\tau^a\pr_i$, is the SU(2) Faddeev-Popov 
operator, generalised to the fundamental representation. Since $FP_f(A)$ is 
linear in $A$, $\Lm$ is easily seen to be convex. Its interior is devoid of 
gauge copies, whereas its boundary $\partial\Lambda$ will in general contain 
gauge copies, associated to vector potentials where the absolute minimum of 
the norm functional are degenerate~\cite{vba3}. It can happen that for some 
points on the boundary the minimum is not quadratic, but of quartic (or higher)
order. The Gribov horizon will touch the boundary of the fundamental domain at 
these so-called singular boundary points, see fig.~1.

It should be noted that the constant gauge degree of freedom is {\em not} 
fixed by the Coulomb gauge condition and therefore one still needs to divide 
by $G$ to get the proper identification, $\Lm/G=\cal{A}/\cal{G}$.
Here $\Lm$ is considered to be the set of absolute minima modulo
the boundary identifications, that remove the degenerate absolute minimum.
It is these boundary identifications that restore the non-trivial topology 
of $\cal{A}/\cal{G}$. There is no problem in dividing out $G$ by
demanding wave functionals to be gauge singlets (colourless states)
with respect to $G$. Because the boundary identifications are by gauge 
transformations, the wave functional will be identified up to a phase factor,
possibly non-trivial when the associated gauge transformation is topologically
non-trivial. The classical scale invariance of the theory guarantees that 
the fundamental domain and the Hamiltonian, when expressed in dimensionless
fields $LA$, only depend on the shape but not on the size of the volume. 
The size dependence will appear solely due to the need of a short distance 
cut-off, giving rise to a scale dependent coupling constant. It is due to
the increase of the effective coupling constant that wave functionals start
to spread out over field space. The modes in which this spreading is largest 
are those associated to transitions over the sphaleron. Non-perturbative 
features become large when the wave functional bites its own tail through the 
boundary identifications. In the available examples this first happens at the 
sphalerons, which lie on the boundary of the fundamental domain and its 
boundary identifications are by gauge transformation with non-trivial topology, 
associated to instantons on whose tunnelling path the sphalerons lie.

\section{Gauge Fields on the Three-Sphere}
The conformal equivalence of $S^3\times\real$ to $\real^4$ allows one to 
construct instantons explicitly~\cite{vbda}. This greatly simplifies the study 
of how to formulate $\theta$ dependence in terms of boundary conditions on 
the fundamental domain~\cite{vbvd}.
We embed $S^3$ in $\real^4$ by considering the unit sphere parametrised by a 
unit vector $n_\mu$. Dependence on the radius $R$ can be retrieved by rescaling
the fields. We introduce $\sg_\mu=(\id,i\vec{\tau})$, which satisfy $\sg_\mu
\sgbar_\nu=\eta^\al_{\mu \nu}\sg_\al$ and $\sgbar_\mu\sg_\nu=\bar{\eta}^\al_{\mu
\nu}\sg_\al$, with $\eta$ the 't Hooft symbols~\cite{thoo}. These can be used to
define orthonormal framings on $S^3$, $e^a_\mu=\eta^a_{\mu\nu}n_\nu$ and $\bar{
e}^a_\mu=\bar{\eta}^a_{\mu \nu}n_\nu$. Note that $e$ and $\bar{e}$ have opposite
orientations. 

\begin{figure}{\tt}
\vspace{5cm}
\includegraphics{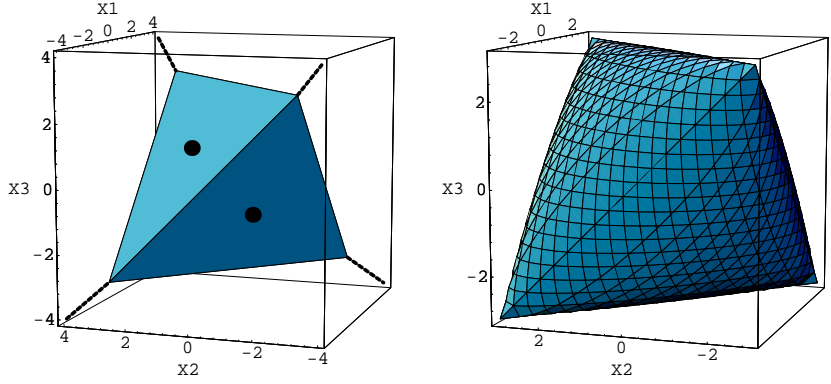}
\caption{
The fundamental domain (left) for constant gauge fields on $S^3$, with respect
to the instanton framing $e_\mu^a$, in the ``diagonal'' representation
$A_a=x_a\sg_a$ (no sum over $a$). By the dots on the faces we indicate the
sphalerons, whereas the dashed lines represent the symmetry axes of the
tetrahedron. To the right we display the Gribov horizon, which encloses
the fundamental domain, coinciding with it at the singular boundary points
along the edges of the tetrahedron.}
\label{fig:fig2}
\end{figure}

The (anti-)instantons in these framings are obtained from those for $\real^4$ 
by identifying the radius in $\real^4$ with the exponential of the time $t$ 
in the space $S^3\times\real$. The (anti-)instanton that tunnels 
through the (anti-)sphaleron, has for each time a constant energy density, 
and is particularly simple with respect to this framing. One finds $A_0=0$,
$A_a=A^\mu e_\mu^a=-f(t)\sigma_a$ for the instanton (and $A_a=A^\mu 
\bar{e}_\mu^a=-f(t)\sigma_a$ for the anti-instanton) with $f(t)=1/(1+e^{-2t})$. 
The (anti-)sphaleron occurs in this parametrisation at $t=0$. It is a saddle 
point of the energy functional with one unstable mode, corresponding to the 
direction of tunnelling. At $t = \infty$, $A_a=-\sg_a$ has zero energy and is 
a gauge copy of $A_a=0$, by a gauge transformation $h=n\cdot\sgbar$ with 
winding number one. This gauge transformation also maps the anti-sphaleron to 
the sphaleron. The two dimensional space containing the tunnelling paths 
through the (anti-)sphalerons is parametrised by $A_\mu(u,v)=\half(-u e^a_\mu-
v\bar{e}^a_\mu)\sg_a$. The gauge transformation $h=n\cdot\sg$ with winding 
number $-1$ is easily seen to map $(u,v)=(w,0)$ into $(u,v)=(0,2-w)$. 
The space of modes degenerate with these and of lowest energy is described by
$A_\mu(c,d)=A_i(c,d)e_\mu^i=\half(c^a_i  e^i_\mu+d^a_j\bar{e}^j_\mu)\sg_a$.
The $c$ and $d$ modes are mutually orthogonal and satisfy the Coulomb gauge 
condition $\pr_i A_i(c,d)=0$. The energy functional is given by~\cite{vbda} 
\bea
  \Ss{V}(c,d)&\equiv&- \int_{S^3} \frac{1}{2} \tr(F_{ij}^2)
  = \Ss{V}(c) + \Ss{V}(d) + \frac{2 \pi^2}{3}
   \left\{ (c^a_i)^2 (d^b_j)^2 - (c^a_i d^a_j)^2 \right\},\nonumber\\
  \Ss{V}(c)&=&2 \pi^2 \left\{ 2 (c^a_i)^2 + 6 \det c + 
  \frac{1}{4}[(c^a_i c^a_i)^2 - (c^a_i c^a_j)^2 ] \right\},
\label{pot}
\eea
from which the degeneracy to second order in $c$ and $d$ can be verified. There
are no modes with a lower zero-point frequency than these~\cite{vbvd}.

An effective Hamiltonian for the $c$ and $d$ modes is derived from 
the one-loop effective action~\cite{vdhe}. To lowest order it is 
given by 
\beq
H=-\frac{g^2(R)}{2R}\left(\left(\frac{\partial}{\partial c_i^a}\right)^2+
\left(\frac{\partial}{\partial c_i^a}\right)^2\right)
+\frac{1}{g^2(R)R}\Ss{V}(c,d),
\eeq
where $g(R)$ is the running coupling constant. It can be shown~\cite{vbvd}
that the boundary of the fundamental domain will touch the Gribov horizon
$\partial\Om$, such that it contains singular points. This is illustrated
in figure 2, which shows the fundamental and Gribov regions for $d=0$, using
the rotational and gauge invariance to rotate $c$ to a ``diagonal'' form.

It is essential that the sphalerons do {\em not} lie on the Gribov horizon and 
that the potential energy near $\partial\Om$ is relatively high as can be seen 
from figure 1. This is why we can take the boundary identifications near the
sphalerons into account without having to worry about singular boundary
points, as long as the energies of the low-lying states will be not much
higher than the energy of the sphaleron. It allows one to study the
glueball spectrum as a function of the CP violating angle $\theta$, but
more importantly it incorporates for $\theta=0$ the noticeable influence
of the barrier crossings, i.e. of the instantons.

\begin{figure}{\tt}
\vspace{5.5cm}
\includegraphics{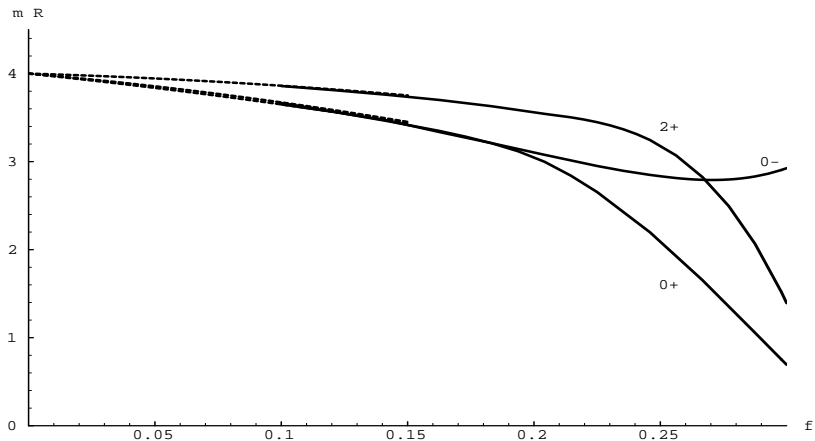}
\caption{The full one-loop results for the masses of scalar, tensor and odd
glueballs on $S^3$ as a function of $f=g^2(R)/2\pi^2$ for $\theta=0$. The
dashed lines correspond to the perturbative result.}
\label{fig:fig3}
\end{figure}

The boundary conditions are chosen so as to coincide with the appropriate 
boundary conditions near the sphalerons, but such that the gauge and (left 
and right) rotational invariances are not destroyed. Projections on the 
irreducible representations of these symmetries turned out to be essential to 
reduce the size of the matrices to be diagonalised in a Rayleigh-Ritz analysis.
Remarkably all this could be implemented in a tractable way~\cite{vdhe}. Results
are summarised in figure 3. One of the most important features is that the 
$0^-$ glueball is (slightly) lighter than the $0^+$ in perturbation theory, 
but when including the effects of the boundary of the fundamental domain, 
setting in at $f\sim 0.2$, the $0^-/0^+$ mass ratio rapidly increases. Beyond 
$f\sim 0.28$ it can be shown that the wave functionals start to feel parts of 
the boundary of the fundamental domain which the present calculation is not 
representing properly~\cite{vdhe}. This value of $f$ corresponds to a 
circumference of roughly 1.3 fm, when setting the scale as for the torus, 
assuming the scalar glueball mass in both geometries at this intermediate 
volume to coincide. 

\section{Conclusion}
Boundary identifications become relevant at large volumes, whereas at very small
volumes the wave functional is localised around $A=0$ and one need not worry 
about these non-perturbative effects. That these effects can be dramatic, even 
at relatively small volumes (above a tenth of a fermi across), was demonstrated
for the case of the torus~\cite{vba1}. Here we have discussed the situation for
$S^3$. Results for the spectrum are compatible with those of a torus in volumes
around one fermi across~\cite{mite}, with $m(2^+)/m(0^+)\sim 1.5$ and
$m(0^-)/m(0^+)\sim 1.7$. For more details and discussions see 
refs.~\cite{vdhe,vba4}. 

\ack{The author thanks the (session) organisers Yitzhak Frishman, Robert 
Perry, Simon Dalley and Pierre Grang\'e for their invitation. He also thanks 
the participants for many fruitful discussions on and off the slopes.}

\end{document}